\documentclass[preprint,showpacs,preprintnumbers,amsmath,amssymb]{revtex4}

\usepackage{graphicx}
\usepackage{bm}

\begin{document}

\title{The Influence of Chemical Short Range Order on Atomic Diffusion in Al--Ni Melts}

\author{S.\ K.\ Das}
\author{J.\ Horbach}\email{horbach@uni-mainz.de}
\affiliation{Institut f\"ur Physik, Johannes Gutenberg--Universit\"at Mainz,
             55099 Mainz, Germany}
\author{M.\ M.\ Koza}
\affiliation{$^4$Institut Laue--Langevin, 38042 Grenoble, France}
\author{S.\ Mavila Chatoth}
\author{A.\ Meyer}\email{ameyer@ph.tum.de}
\affiliation{Physik Department E\,13, Technische Universit\"at M\"unchen, 85747
Garching, Germany}

\date{\today}

\begin{abstract}
We use inelastic neutron scattering and molecular dynamics (MD) simulation
to investigate the chemical short range order (CSRO), visible through
prepeaks in the structure factors, and its relation to self diffusion in
Al--Ni melts.  As a function of composition at 1795\,K Ni self diffusion
coefficients from experiment and simulation exhibit a non-linear
dependence with a pronounced increase on the Al--rich side. This comes
along with a change in CSRO with increasing Al content that is related
to a more dense packing of the atoms in Ni--rich Al--Ni systems.
\end{abstract}


\maketitle
In hard-sphere like metallic liquids mass transport is strongly connected
to the packing fraction of the atoms \cite{PrAP73,FoGS03,MaMK04}.
However, in liquid Al--Ni \cite{maret90}, Al--Fe \cite{JiXS98,IlSS02}
and glass forming Al rich alloys \cite{LiYX99} the structure clearly
deviates from a random hard-sphere packing and exhibits a chemical
short range order (CSRO) \cite{egami}.  As was shown in a neutron
diffraction experiment on liquid Al$_{80}$Ni$_{20}$ \cite{maret90}, CSRO
manifests itself in a prepeak in the static partial structure factors
at an intermediate lengthscale around $1.8$~\AA$^{-1}$.  In this paper,
we study the composition dependence of the CSRO in Al--Ni melts and
show how it affects the diffusion dynamics in this alloy.  To address
this issue, we performed inelastic neutron scattering experiments and
molecular dynamics (MD) simulations.

Al--Ni alloys were prepared by arc melting of pure elements
under a purified Argon atmosphere.  The corresponding solidus and
liquidus temperatures were measured with differential scanning
calorimetry and found to be in excellent agreement with the phase
diagram~\cite{predel}. For the neutron time--of--flight experiment
on the IN\,6 spectrometer of the Institut Laue-Langevin, thin--walled
Al$_2$O$_3$ containers were used that provide a hollow cylindrical
sample geometry of 22\,mm in diameter and a thickness of 1.2\,mm for
the Al rich samples, and 0.6\,mm for the Al$_{25}$Ni$_{75}$ sample.
An incident neutron wavelength of $\lambda\!=\!5.1\,\mbox{\AA}$
yielded an energy resolution of $\delta E\simeq92\,\mu\mbox{eV}$
(FWHM) and an accessible wave number range at zero energy transfer
of $q=0.4-2.0\,\mbox{\AA}^{-1}$.  Al$_{100-x}$Ni$_{x}$ alloys were
measured above their liquidus temperatures at 1525\,K and at 1795\,K
for $x=10,20,23,30$, and at 1795\,K for $x=38,75$ compositions.

The scattering law $S(q,\omega)$ was obtained by normalization to
a vanadium standard, correction for self absorption and container
scattering, and interpolation to constant wave numbers $q$.  Further,
$S(q,\omega)$ was symmetrized with respect to the energy transfer
$\hbar\omega$ by means of the detailed balance factor.  Fourier
transformation of $S(q,\omega)$, deconvolution of the instrumental
resolution, and normalization with the value at $t=0$ gave the time
correlation function $\Phi(q,t)$.  For times above $\simeq 1$\,ps,
$\Phi(q,t)$ shows an $\exp(-q^2 D_q t)$ decay to zero.  Towards small
$q$ incoherent scattering on the Ni atoms dominates the signal and
the $q$-dependent diffusivity $D_q$ becomes constant.  Thus, Ni self
diffusion coefficients $D$ could be derived on an absolute scale
\cite{MaMK04,Mey02}.

In order to model the interaction between the atoms in the MD simulation,
we used a potential of the embedded atom type that was recently derived
by Mishin {\it et al.}~\cite{mishin03}. The simulations were done at
$1795$~K and $1525$~K for different Al--Ni compositions.  The systems
consist of 1500 particles in each case.  First standard Monte--Carlo (MC)
simulations in the $NpT$ ensemble~\cite{binder_book} were used to fully
equilibrate the systems at zero pressure and to generate five independent
configurations for MD simulations in the microcanonical ensemble.
In the latter case, Newton's equations of motion were integrated with
the velocity Verlet algorithm using a time step of 1.0~fs. From the
MD trajectories the structure factors and diffusion constants were
determined~\cite{binder04}.

By integrating the scattering law $S(q,\omega)$ over the quasielastic line
inelastic neutron scattering also gives access to structural information on
intermediate length scales.  Fig.~1 displays 
the integral over the quasielastic line (denoted by $S_{\rm qu}(q)$ in the following)
for the Al-Ni melts at $1795$\,K
For clarity a constant that approximates the incoherent scattering
on the Ni atoms has been subtracted.  For Al-rich alloys the spectra
exhibit a prepeak that is increasing with increasing Ni content and that
shifts from $\simeq 1.6$\,\AA$^{-1}$ in Al$_{80}$Ni$_{20}$ to $\simeq
1.8$~\AA$^{-1}$ in Al$_{62}$Ni$_{38}$.  In Al$_{90}$Ni$_{10}$ nearly
no prepeak is visible anymore.  As compared to the data at 1795\,K, the
quasielastic structure factors measured at 1525\,K do not reveal a measurable 
shift in the peak positions and the intensities in the maxima of the prepeaks
exhibit a $\simeq 20$\,\% increase which can be accounted for by the
increased Debye-Waller factor.  On the Ni rich side the Al$_{25}$Ni$_{75}$
$S_{\rm qu}(q)$ also exhibits a prepeak.  This prepeak is much broader,
though, and has its maximum at $\simeq 1.5$\,\AA$^{-1}$.

Fig.~2 shows the measured Ni self diffusion coefficients as
a function of composition at 1525\,K and at 1795\,K.  At 1795\,K
values range from $3.95\pm0.10\times10^{-9}$m$^2$s$^{-1}$ in
Al$_{25}$Ni$_{75}$ to $10.05\pm0.11\times10^{-9}$m$^2$s$^{-1}$
in Al$_{90}$Ni$_{10}$.  The Ni self diffusion coefficient in
Al$_{25}$Ni$_{75}$ is equal within error bars to the value in pure
liquid Ni with $D=3.80\pm0.06\times10^{-9}$m$^2$s$^{-1}$ \cite{MaMK04}.
Substitution of 25\,\% Ni by Al does not affect the diffusion coefficient
significantly.  In contrast, on the Al rich side diffusion coefficients
show a pronounced increase with increasing Al content.

The temperature and concentration dependence of the measured diffusion
coefficients are well reproduced by the ones of the MD simulation: There
is an overall 20\% agreement with the experimental data 
(Fig.~2). From the inset of Fig.~2 we see also that in our simulation
the ratio $D_{\rm Al}^*/D_{\rm Ni}^*$ does not depend on temperature in
the considered temperature range (here the star
means that diffusion constants from the
simulation are considered).  Note that the diffusion constants,
that were obtained in a recent simulation study by Asta {\it et
al.}~\cite{asta99} using different EAM potentials, are a factor 2--3
higher than those obtained in our simulation, but also display a
pronounced non--linear behavior as a function of composition.  As we
shall see in the following, the non--linear behavior of the diffusion
coefficients is reflected in structural changes that can be understood
from the detailed informatiom provided by the simulation.

Fig.~3 displays the 
$
S_{\rm n}(q)=\frac{1}{N_{\rm Al}b_{\rm Al}^2+N_{\rm Ni}b_{\rm Ni}^2}
\sum_{k,l}^N b_k b_l \langle {\rm e}^{i{\bf q}\cdot
({\bf r}_k -{\bf r}_l)} \rangle,
$
with ${\bf r}_k$ being the position of the $k$'th particle.  $N_{\alpha}$,
$\alpha \in {\rm Al, Ni}$, denotes the number of atoms of type $\alpha$,
and $b_{\rm Al}=0.3449\cdot 10^{-12}$\,cm and $b_{\rm Ni}=1.03\cdot
10^{-12}$\,cm are respectively the neutron scattering lengths of Al
and Ni.  As Fig.~3 shows, also a prepeak is observed in the
simulation which is located at slightly smaller $q$ for the Al rich
systems whereas for Al$_{30}$Ni$_{70}$ the prepeak is at a slightly
larger $q$ than expected from the experiment. However, in particular for
the Al rich systems a similar behavior as in the experiment is found,
i.e., with increasing Al concentration, the prepeak becomes broader,
its amplitude decreases and its location shifts to smaller $q$.  We note
that $S_{\rm qu}(q)$ and $S_{\rm n}(q)$ are {\it different} quantities,
since $S_{\rm qu}(q)$ is also affected by the $q$ dependence of the
Debye--Waller factor. Moreover, $S_{\rm qu}(q)$ contains a mixture of
incoherent and coherent contributions that cannot be disentangled from
each other.  Thus, one should compare only the {\it relative} intensities
of the peaks in the experimental $S_{\rm qu}(q)$.

Also shown in Fig.~3 is the main peak which indicates repeating
units of neighboring atoms.  The location of this peak moves to higher
$q$ towards the Ni rich systems which is due to the fact that with
increasing Ni concentration the packing density of the atoms increases
\cite{remark}.  This can be inferred from the inset of Fig.~3
which shows the atomic volume $V_{\rm a}$ as found in the simulation in
comparison to experimental data~\cite{ayushina69}. Note that we get an
overall 5\% agreement with the experiment over the whole Al concentration
range. The nonlinear behavior of $V_{\rm a}$ is strongly correlated with
the nonlinear behavior of the diffusion constants as a result of the CSRO.

$S_{\rm qu}(q)$ and $S_{\rm n}(q)$ are total
structure factors which are in binary systems 
linear combinations of three partial structure
factors. In order to understand the CSRO that is present in
Al--Ni melts one has to consider these partial structure
factors defined by
%
$
   S_{\alpha \beta}(q)=
       \frac{1}{\sqrt{N_{\alpha} N_{\beta}}}
                \sum_{k_{\alpha}=1}^{N_{\alpha}} 
                \sum_{l_{\beta}=1}^{N_{\beta}}
          \langle 
          \exp(i{\bf q}\cdot ({\bf r}_{k_{\alpha}} -{\bf r}_{l_{\beta}} )) 
         \rangle,
$
%
where the indices $k_{\alpha}$ and $l_{\beta}$ correspond to particles
of type $\alpha$ and $\beta$, respectively. In Fig.~4
the different $S_{\alpha \beta}(q)$ are shown again at $T=1525$~K
for Al rich compositions, for a 50:50 mixture, and for the Ni rich
composition Al$_{30}$Ni$_{70}$.  The emergence of a prepeak in
$S_{\alpha \beta}(q)$ indicates that there are repeating structural
units involving next--nearest $\alpha \beta$ neighbors which are built
in inhomogeneously into the structure.  As we see in Fig.~4,
no prepeak emerges in $S_{\rm AlAl}(q)$ for the Al rich compositions
but the main peak moves slightly to higher $q$ with decreasing Al
concentration. By further decreasing the Al concentration a prepeak
evolves that moves to smaller $q$. As a result a double peak structure
is observed in $S_{\rm AlAl}(q)$ for Al$_{30}$Ni$_{70}$ with maxima
at $q_1=1.8$~\AA$^{-1}$ and $q_2=3.0$~\AA$^{-1}$. In $S_{\rm AlNi}(q)$
a prepeak with a negative amplitude is visible at all Al concentration
and, as in the Al--Al correlations, this prepeak moves to smaller $q$
with decreasing Al concentration.  The negative amplitude
of the prepeak in $S_{\rm AlNi}(q)$ is due to the avoidance of
corresponding Al--Ni distances. Also in $S_{\rm NiNi}(q)$ a prepeak is
present at all concentrations.  This prepeak broadens with increasing
Al concentration while its location remains essentially constant. 
In Al$_{30}$Ni$_{70}$ the main peak is at the same
location in the three different partial structure factors, namely around
$q=3.0$~\AA$^{-1}$.

The picture provided by the simulation is as follows: In the Ni rich
systems the minority species Al is built into the structure such that
the distance between repeated Al--Al units corresponds to that of
repeated Ni--Ni units~\cite{remark} whereas towards the Al rich systems
the distance between repeated Al--Al units becomes larger without a
significant change in the distance between repeated Ni--Ni units. All
this explains the nonlinear behavior of the atomic volume and, since
it indicates a nonlinear behavior of the packing of the atoms upon a
change of the composition, it demonstrates that the nonlinear increase
of the diffusion constants upon increasing Al content is essentially a
packing effect. Of course, this packing argument does not explain
the deviation of $D_{\rm Al}^*/D_{\rm Ni}^*$ from one for the Ni rich systems.
The understanding of this feature requires a microscopic theory.

In conclusion, an investigation of Al--Ni melts using inelastic neutron
scattering and MD simulation has shed light onto the nonlinear composition
dependence of the diffusion constants and its relation to chemical short
range order visible through a prepeak in the structure factor.  We confirm
the finding for other hard-sphere like metallic melts~\cite{MaMK04}
that the transport in these systems is strongly affected by packing
effects although the interactions between the atoms depend strongly on
chemical details.

We are grateful to Kurt Binder for stimulating discussion.  The authors
acknowledge financial support by the Deutsche Forschungsgemeinschaft
(SPP Phasenumwandlungen in mehrkomponentigen Schmelzen) under Grant
No. Bi314/18 and Me1958/2. One of the authors (J. H.) acknowledges the
support of the DFG under Grant No. HO 2231/2.

\newpage

\noindent
FIG.~1: Quasielastic coherent structure factor \protect$S_{\rm qu}(q)$ at intermediate $q$ 
of Al--Ni melts at 1795\,K as obtained by inelastic neutron scattering (see
text).\\ 

\noindent
FIG.~2: Diffusion coefficients at 1795\,K and at 1525\,K as a function of the aluminium
concentration from simulation and neutron scattering. The broken lines are guides 
to the eye. Inset: Ratio of Al and Ni self diffusion coefficients from simulation.
Note that the diffusion constants from the simulation are decorated by a star.\\

\noindent
FIG.~3: Structure factors \protect$S_{\rm n}(q)$ of Al--Ni melts as calculated from the
partial structure factors of the simulation weighted with the neutron scattering length.
Inset: Atomic volume from the simulation compared to experimental data from
Ref.~\protect\cite{ayushina69} as a function of composition.\\

\noindent
FIG.~4: Partial structure factors \protect$S_{\alpha \beta}(q)$ of Al--Ni melts as determined
by the simulation at 1525~K.  a) \protect$S_{\rm AlAl}(q)$, b) \protect$S_{\rm AlNi}(q)$, 
and c) \protect$S_{\rm NiNi}(q)$.\\

\newpage
\vspace*{0.5cm}
\begin{figure}[t]
\includegraphics[width=132mm]{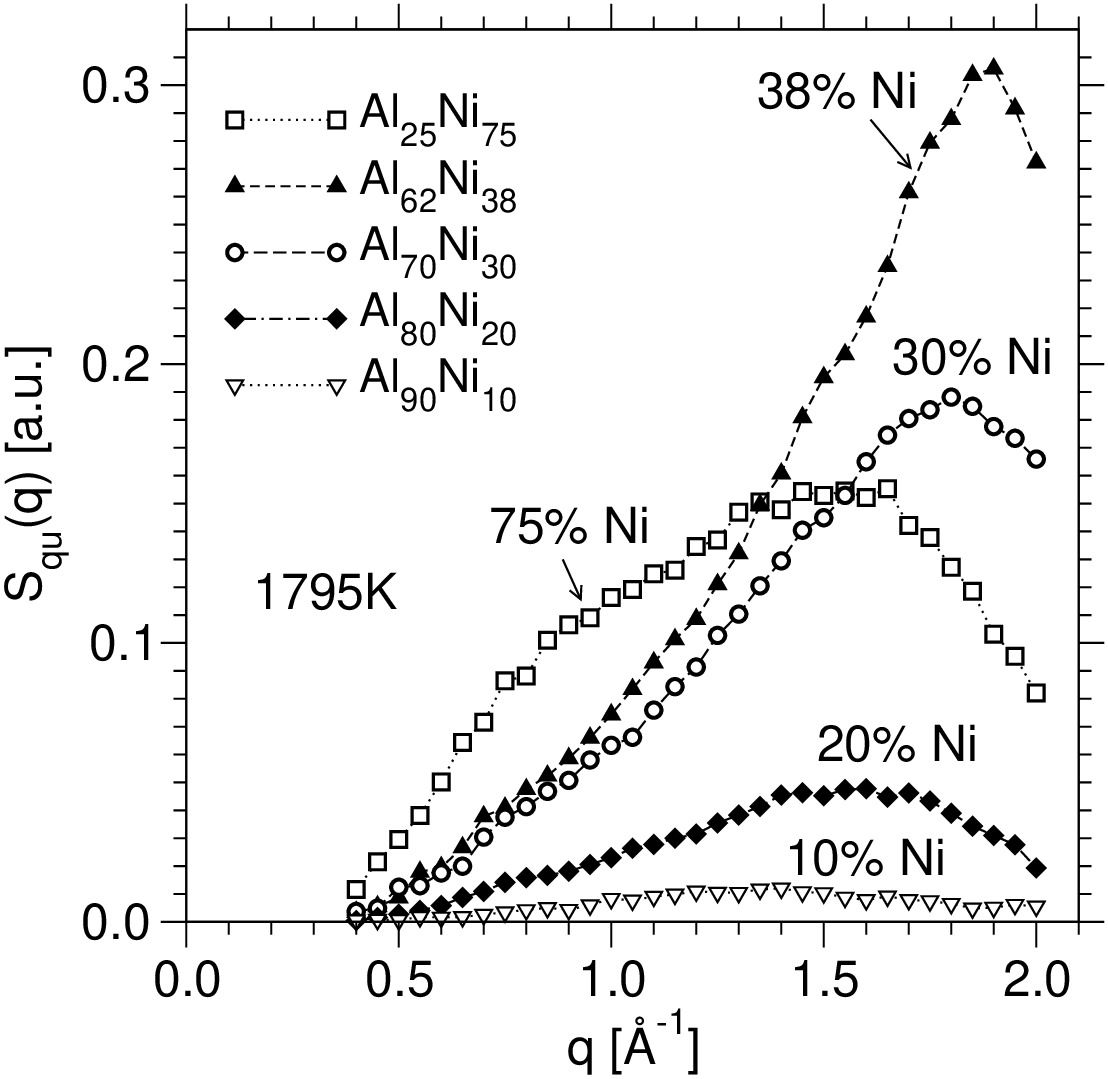}
\\
\vspace*{0.5cm}
\hspace*{-1.85cm}{\LARGE Das et al., Figure 1}
\end{figure}

\newpage
\vspace*{0.5cm}
\begin{figure}[t]
\includegraphics[width=132mm]{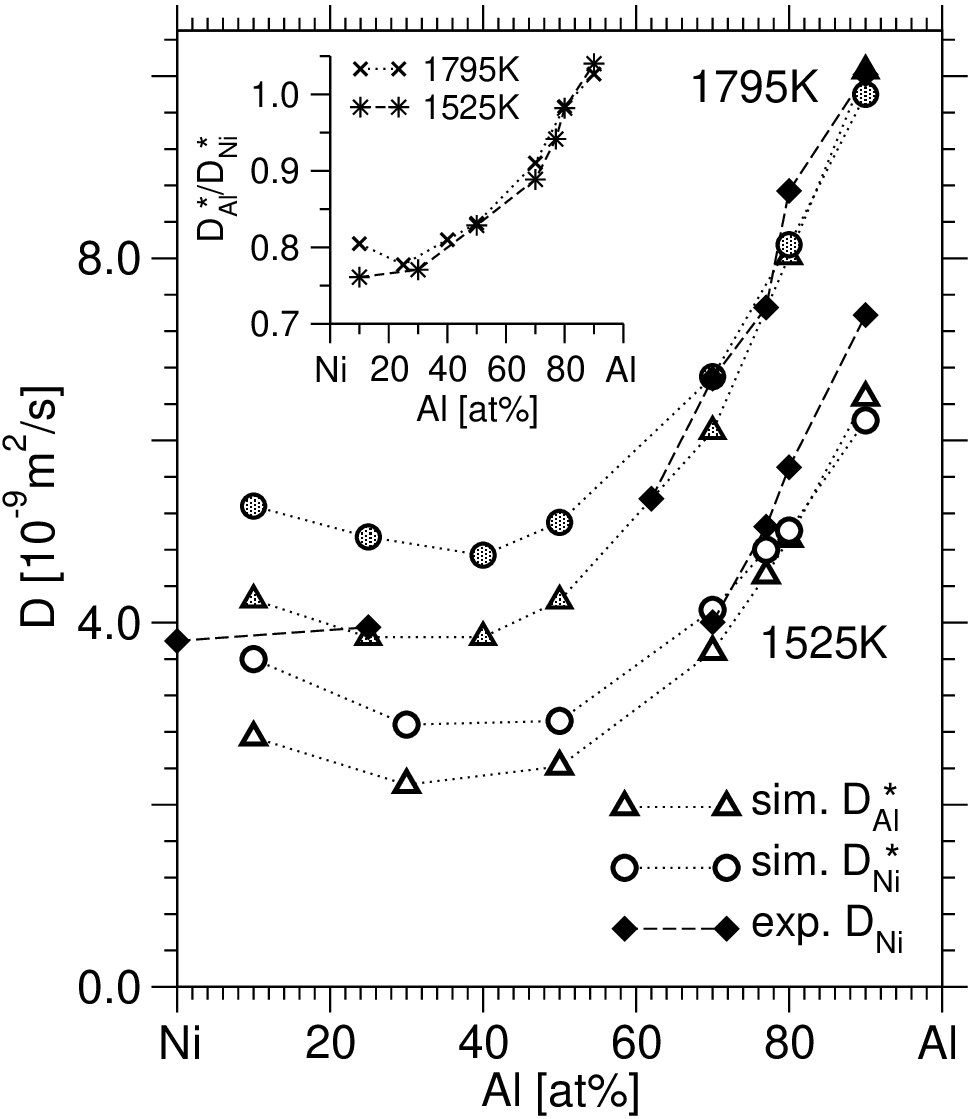}
\\
\vspace*{0.5cm}
\hspace*{-1.55cm}{\LARGE Das et al., Figure 2}
\hspace*{1cm}{\,}
\end{figure}

\newpage
\vspace*{0.5cm}
\begin{figure}[t]
\includegraphics[width=132mm]{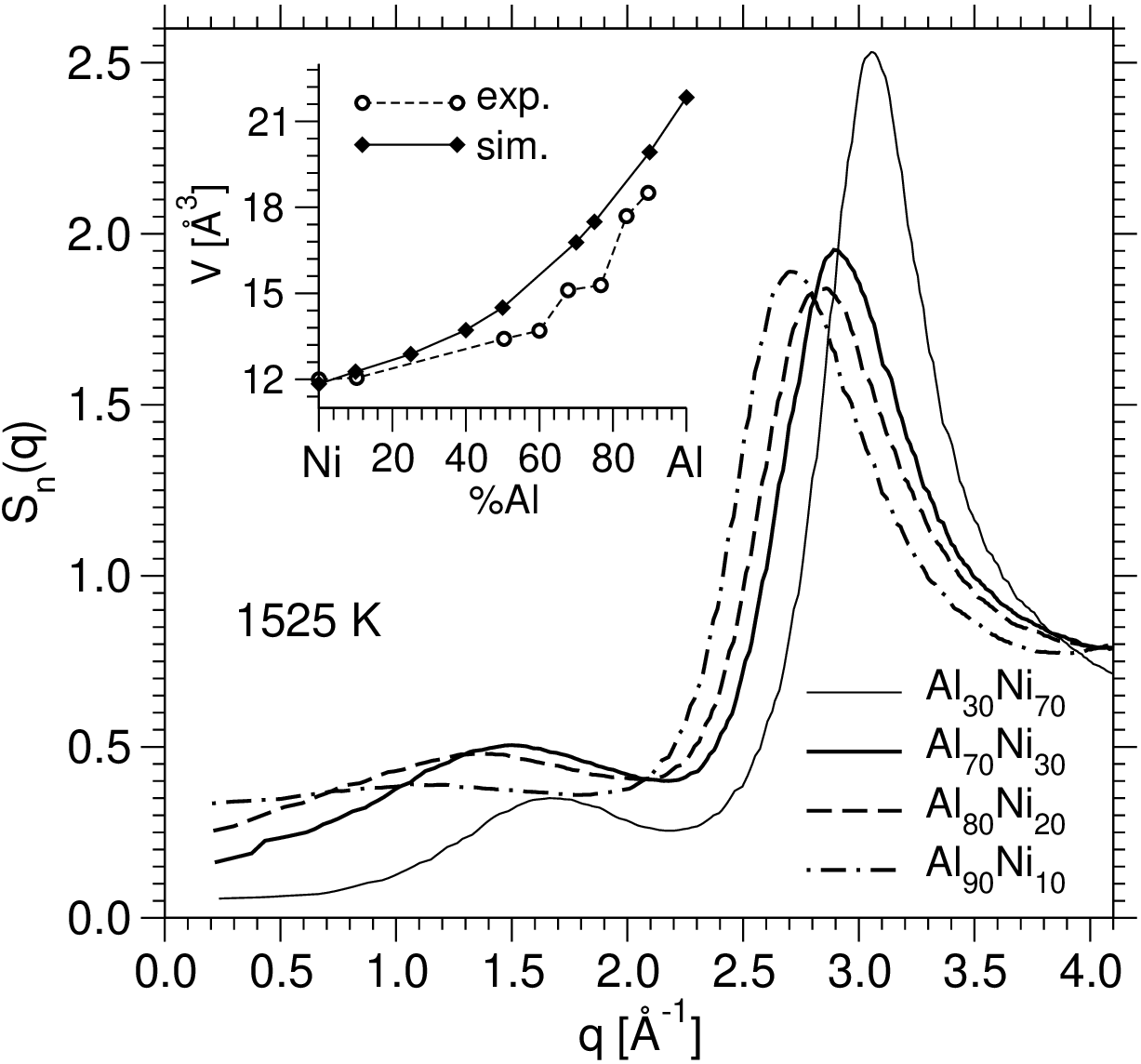}
\\
\vspace*{0.5cm}
\hspace*{-1.75cm}{\LARGE Das et al., Figure 3}
\hspace*{1cm}{\,}
\end{figure}

\newpage
\vspace*{0.5cm}
\begin{figure}[t]
\includegraphics[width=132mm]{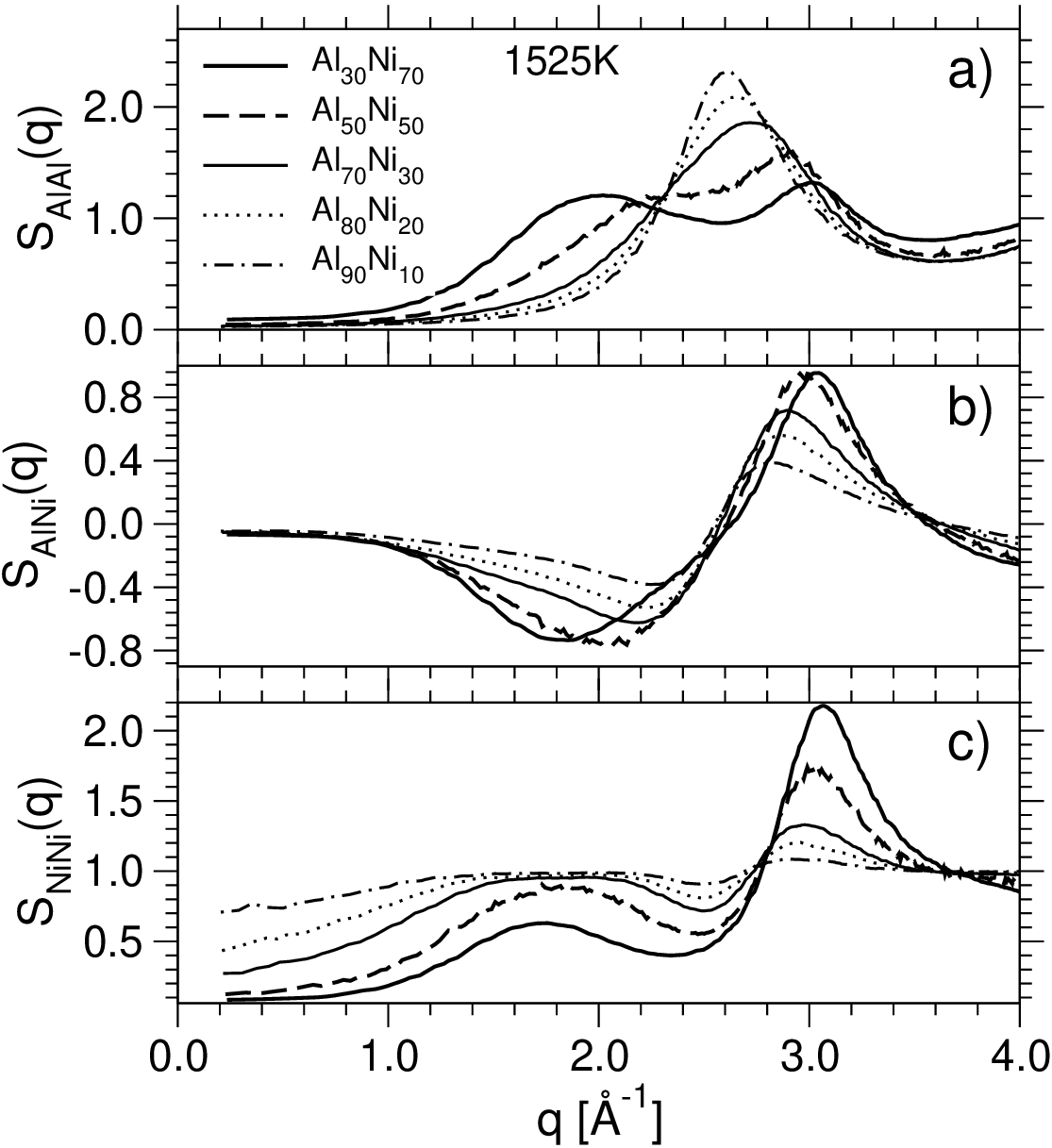}
\\
\vspace*{0.5cm}
\hspace*{-1.65cm}{\LARGE Das et al., Figure 4}
\hspace*{1cm}{\,}
\end{figure}


\newpage

\begin{thebibliography}{1}

\bibitem{PrAP73} P. Protopapas, H. C. Andersen, and N. A. D. Parlee, J. Chem. Phys.
                 {\bf 59}, 15 (1973).
\bibitem{FoGS03} G. Foffi, W. G\"otze, F. Sciortino, P. Tartaglia, and Th. Voigtmann, 
                 Phys. Rev. Lett. {\bf 91}, 085701 (2003).
\bibitem{MaMK04} S. Mavila Chathoth, A. Meyer, M. M. Koza, and F. Yuranji, Appl. Phys. Lett.
                 (in press).
\bibitem{maret90} M. Maret, T. Pomme, A. Pasturel, and P. Chieux, 
                  Phys. Rev. B {\bf 42}, 1598 (1990).
\bibitem{JiXS98} Q. Jingyu, B. Xiufang, S. I. Sliusarenko, and W. Weimin,
                 J. Phys.: Condens. Matter {\bf 10}, 1211 (1998).
\bibitem{IlSS02} A. Il'inskii, S. Slyusarenko, O. Slukhovskii, I. Kaban, and W. Hoyer,
                 Mater.\ Sci.\ Engin.\ A {\bf 325}, 98 (2002).
\bibitem{LiYX99} Z. Lin, W.\ Youshi, B. Xiufang, L. Hui, W. Weimin,
                 L. Jingguo, and L. Ning, 
                 J. Phys. Condens. Matter {\bf 11}, 7959 (1999).
\bibitem{egami} T. Egami in F. E. Luborsky (ed.), {\it Amorphous Metallic Alloys} 
                (Butterworths, London, 1983).
\bibitem{predel} B. Predel, edited by O. Madelung, 
                 {\it Phase Equilibria of Binary Alloys} (Springer, Berlin, 2003).
\bibitem{Mey02} A. Meyer, Phys. Rev. B {\bf 66}, 134205 (2002).
\bibitem{mishin03} Y. Mishin, M. J. Mehl, and D. A. Papaconstantopoulos,
                   Phys. Rev. B {\bf 65}, 224114 (2002). 
\bibitem{binder_book} D. P. Landau and K. Binder,
                 {\it A Guide to Monte Carlo Simulations in Statistical Physics}
                 (Cambridge University Press, Cambridge, 2000).
\bibitem{binder04} K. Binder, J. Horbach, W. Kob, W. Paul, and F. Varnik,
                   J. Phys.: Condens. Matter {\bf 16}, S429 (2004).
\bibitem{asta99} M. Asta, D. Morgan, J. J. Hoyt, B. Sadigh, J. D. Althoff, D. de Fontaine, 
                 and S. M. Foiles, Phys. Rev. B {\bf 59}, 14271 (1999). 
\bibitem{ayushina69} G. D. Ayushina, E. S. Levin, and P. V. Geld,
                     Russ. J. Phys. Chem. {\bf 43}, 2756 (1969).
\bibitem{remark} The location of the first peak in the corresponding
                 pair correlation functions $g_{\alpha \beta}(r)$ 
                 does not depend on the Al--Ni composition.

\end{thebibliography}
\end{document}